# On the Role of Crystal Defects on the Lattice Thermal Conductivity of Monolayer WSe$_2$ (P63/mmc) Thermoelectric Materials by DFT Calculation

Yingtao Wang,[1] Xian Zhang[1,*]

*1 Department of Mechanical Engineering, Stevens Institute of Technology, 1 Castle Point Terrace, Hoboken, NJ 07030, USA*

**Abstract**

As the energy problem becomes more prominent, research on thermoelectric (TE) materials has deepened over the past few decades. Low thermal conductivity enables thermoelectric materials better thermal conversion performance. In this study, based on the first principles and phonon Boltzmann transport equation, we studied the thermal conductivities of single-layer WSe$_2$ under several defect conditions using density functional theory (DFT) as implemented in the Vienna Ab-initio Simulation Package (VASP). The lattice thermal conductivities of WSe$_2$ under six kinds of defect states, i.e., PS, SS-c, DS-s, SW-c, SS-e, and DS-d, are 66.1, 41.2, 39.4, 8.8, 42.1, and 38.4 W/(m·K), respectively at 300 K. Defect structures can reduce thermal conductivity up to 86.7% (SW-c) compared with perfect structure. The influences of defect content, type, location factors on thermal properties have been discussed in this research. By introducing atom defects, we can reduce and regulate the thermal property of WSe$_2$, which should provide an interesting idea for other thermoelectric materials to gain a lower thermal conductivity.

## 1. Introduction

With the shortage and quality degradation of resources becoming increasingly prominent, the development of new sustainable green energy has become an important issue [1, 2]. In the industrial production process, a large amount of waste heat will be generated and discharged into the environment, which not only causes waste, but may also cause thermal pollution [3, 4]. It is of great significance to the recycling and reuse of waste heat. The 2D materials family grows rapidly since the successful synthesis of graphene in 2004 [5] due to their outstanding chemical and physical property [6, 7]. Up to now, they have applied to numerous fields, involving energy storage and conversion [8, 9], catalysis [10, 11], electron device [12], photoelectric device [13], medical treatment [14], etc.

TE materials can be applied to plenty of devices, which transform waste heat into electrical energy [15]. The dimensionless figure of merit, ZT, is one critical property for TE materials to determine the energy conversion efficiency. Besides high ZT values, prominent TE materials should have low thermal conductivity, $\kappa$ (enable better energy conversion performance) and high electrical conductivity, $\sigma$ (enable better electrical conduction) [16, 17]. It is challenging to have both simultaneously since these two properties are interrelated and determined by the Wiedemann-Franz Law.

WSe$_2$, one kind of 2D transition metal dichalcogenides (TMDCs), has low lattice thermal conductivity, thus has gained a lot of attention [18, 19]. Unlike abundant studies on graphene and MoS$_2$, the systematic and theoretical research on thermal transport properties of WSe$_2$ is still in its primary stage [20]. Therefore, systematic and targeted theoretical research is of great significance for understanding the heat transfer mechanism of WSe$_2$, understanding its thermoelectric conversion principles, and guiding experiments and future commercial applications. Thermal conductivity is a very important parameter in many applications. Mechanical strain is an effective and widely used method to adjust the thermal conductivity of materials and has been studied intensively [21, 22]. Through the delicate design, the thermal conductivity can be greatly changed by applying a small mechanical strain. Also, by constructing heterostructures, the thermal conductivity can be effectively adjusted [23, 24]. By using classical non-equilibrium molecular dynamics (NEMD) simulations, Rahman et.al studied the phonon thermal conductivity of stanine/hBN, found that the bulk thermal conductivity at room temperature is 15.20 W/(mK), 550 W/(mK), and 232 W/(mK) for bare stanene and hBN, and stanene/hBN, respectively [24]. By introducing defects is another way to adjust the thermal conductivity. In this work, based on the first principles and phonon Boltzmann transport equation, this research will study the thermal conductivities of single-layer WSe$_2$ under various defect conditions using density functional theory (DFT) as built in VASP.

## 2. Material and methods

VASP installed in high-performance supercomputer clusters was used to perform all first principle calculations with DFT as the basis. The interaction between electrons and ions was described by the projector augmented wave (PAW) pseudopotentials. The generalized gradient approximation (GGA) of the Perdew-Burke-Ernzerhof (PBE) form was used to explain the exchange and correlation interactions between electrons. 400 eV energy cutoff was selected to characterise the Kohn-Sham wavefunctions. The first Brillouin zone was divided by $4 \times 4 \times 4$ k-mesh. All associated structures could be fully relaxed until the energy convergence conditions of $10^{-5}$ eV was fulfilled.

By solving phonon Boltzmann transport equations (BTE), the lattice thermal conductivity could be calculated by the ShengBTE software [25]. The second-order harmonic (can be calculated by VASP package and extracted by phonopy package [26]) and third-order anharmonic interatomic force constants (IFCs) (can be calculated by VASP package and summarized by thirdorder package using cutoff distance of 0.4 nm [27]) are necessary as inputs files. Finally, the lattice thermal conductivity can be calculated by ShengBTE process with a well-converged $5 \times 5 \times 5$ $q$-grid.

## 3. Results and discussions

First, the crystal structure of unit cell WSe$_2$ has been optimized, with the lattice constant being a=3.32524 Å. WSe$_2$ has a hexagonal lattice structure with stacked Se-W-Se layers. The W atom layer is sandwiched between two Se atom layers. Fig. 1a shows the top and side views of the atomic structure of monolayer WSe$_2$, respectively. Referring Brillouin zones is shown in Fig. 1b. It can be seen from the band structure as shown in Fig.1c, WSe$_2$ is a direct semi-conductor with a gap of 2 eV.

Phonon scattering has a very important influence on the heat transport properties of materials. Figure 2 shows the phonon scattering spectrum of a perfectly structured single-layer WSe$_2$ material along a highly symmetrical path Γ-M-K-Γ. A WSe$_2$ unit cell is composed of three atoms, so the spectrum is composed of three acoustic and six optical branches. There is a clear frequency gap between the acoustic and optical branches. This is due to the difference in relative atomic mass between W and Se. This gap is crucial for the heat transport properties of the crystal. The larger the gap, the higher the thermal conductivity.

To understand the effect of defects, defect content, type, and location on the thermal conductivity of WSe$_2$, we set some kinds of defect conditions. The six defect structures are shown in Figure 3 and 4. These structures can be divided into several groups for discussion. Figure 3a shows the crystal structure of the perfect structure, which is labelled PS. Figure 3b shows the crystal structure of a Se defect located in the central area, which is marked SS-c (single Se-center). Figure 3c shows the crystal structure of two Se vacancies located in the same layer, labelled as DS-s (double Se-same layer). Figure 3d shows a crystal structure of the W atom vacancy in the center part, which is identified as SW-c (single W-center). Figure 5a shows the crystal structure of a Se defect located at the edge, which is characterized as SS-e (single Se-edge). Figure 5b shows the crystal structure of two Se atom vacancies located in different layers, marked as DS-d (double Se-different layers). By comparing PS, SS-c, DS-s, the effect of defect content on thermal conductivity can be discussed. Through assessing PS, SS-c, SW-c, the influence of defect atom types on thermal conductivity will be revealed. Finally, by contrasting SS-c and SS-e, DS-s and DS-d, the impact of defect location on thermal conductivity will be explained.

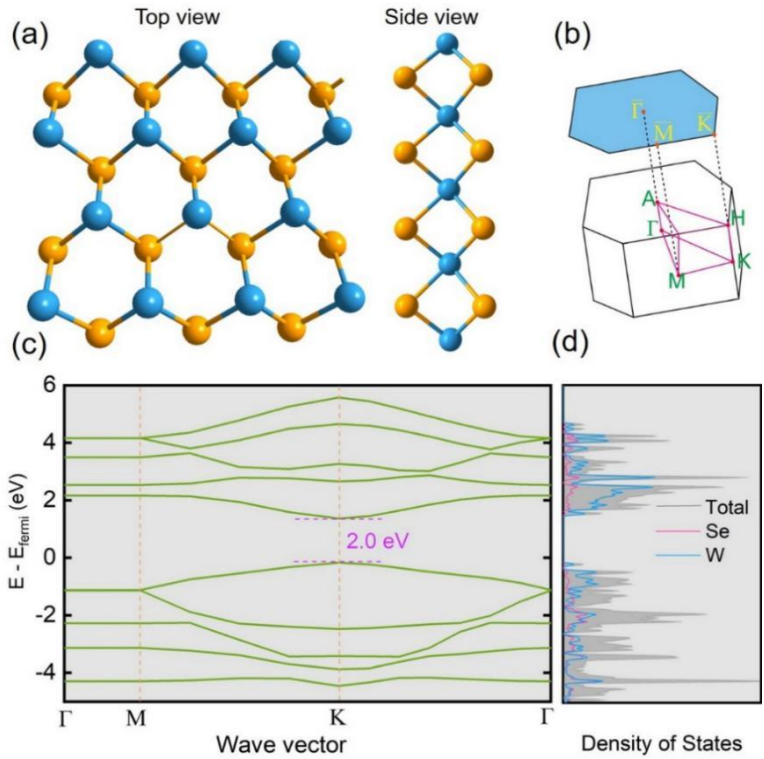

Figure 1. a) Schematic diagram, b) first Brillouin zone, c) band structure, d) density of states of perfect WSe$_2$ structure.

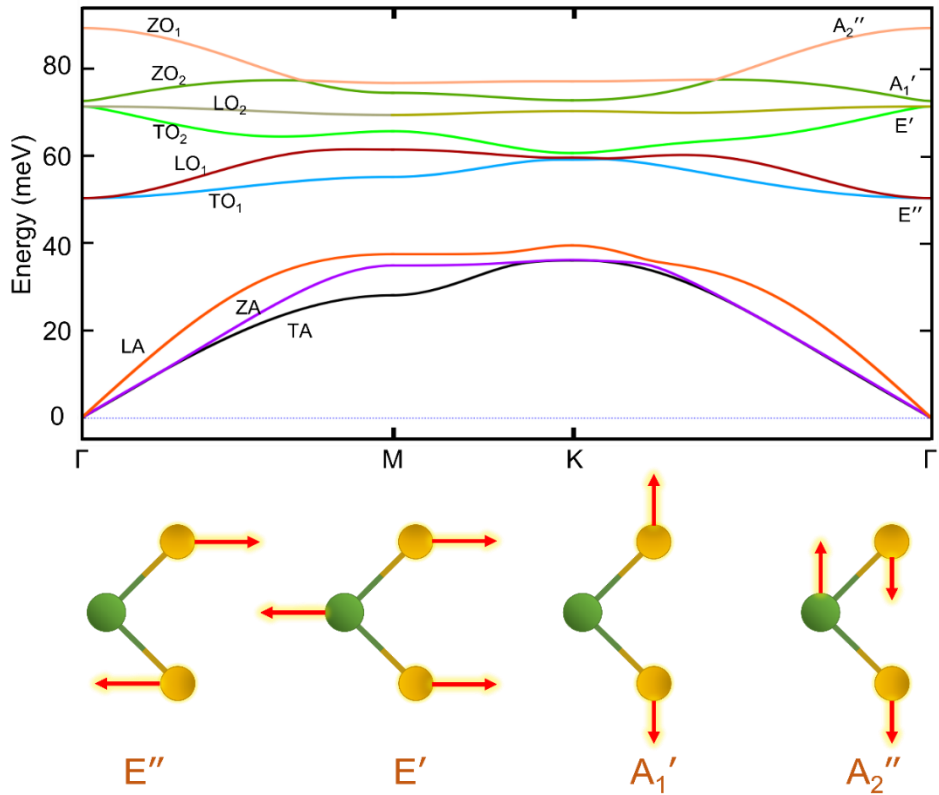

Figure 2. Phonon dispersion spectrum and vibrational modes of PS.

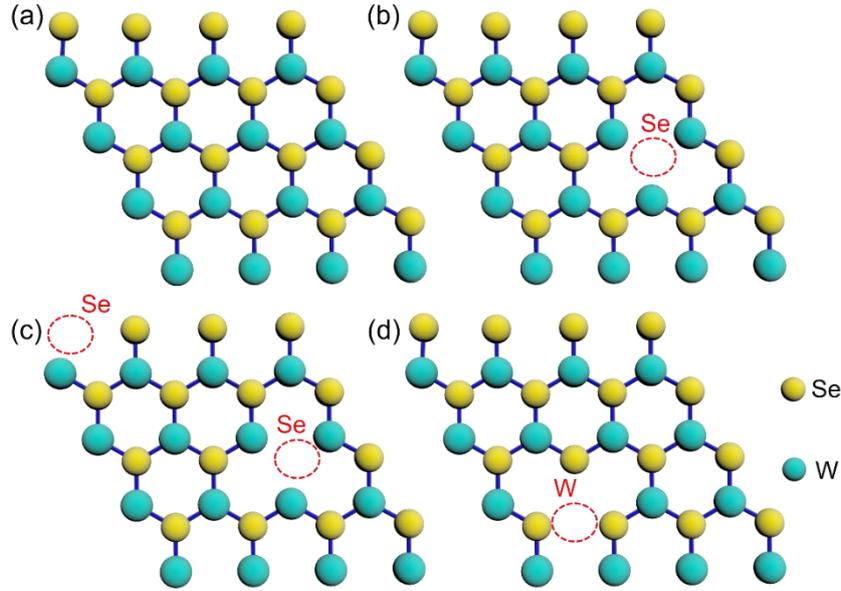

Figure 3. Schematic diagram of defect structures. (a) PS. (b) SS-c. (c) DS-s. (d) SW-c.

The temperature dependence of phonon thermal conductivity in monolayer WSe$_2$ is shown in Fig. 4. The lattice thermal conductivity of PS at 300 K is 66.1 W/(m·K) (Table 1), basically consistent with the values and trends in the literature [28, 29], which can confirm the effectiveness of the method. The thermal conductivities of WSe$_2$ at 300 K are 66.1, 41.2, 39.4, 8.8, 42.1, and 38.4 W/(m·K), respectively for PS, SS-c, DS-s, SW-c, SS-e and DS-d. Sample with defects have much lower thermal conductivity compared with perfect structure, up to 87% decrease (SW-c) [30].

The content of defects has an important influence on thermal conductivity. By comparing the thermal conductivity of the three structures of PS, SS-c, and DS-s (Figure 4, Table 1), it is found that the existence of defects will reduce the thermal conductivity, and the thermal conductivity of crystal structure of two vacancies of Se atoms can be lower than the structure of one vacancy [31]. For instance, at 300 K, the thermal conductivities of the three structures of PS, SS-c, and DS-s are 66.1, 41.2 and 39.8 W/(m·K), respectively. Single-atom defects can reduce the thermal conductivity to 62.3%, and diatomic defects can reduce the thermal conductivity to 60.2%. Due to the introduction of defects, the phonon-defects scattering mechanism leads to the reduction of phonon transmission coefficient [32]. The existence of defects causes the crystal structure gradually become amorphous, destroys the original phonon transport behaviours, renders the heat transport inefficient, and decreases the thermal conductivity [32].

The types of defects also influence thermal transport property [33]. By comparing the thermal conductivity of PS, SS-c, SW-c, it is found that W atom defects allow the structure to obtain much lower thermal conductivity, and even a single W atom defect create a quite lower thermal conductivity than double Se atom defect structure. For instance, at 300 K, the thermal conductivity of the PS, SS-c, and SW-c structures are 66.1, 41.2, and 8.8 W/(m·K), respectively. Monoatomic W defects can reduce the thermal conductivity by 86.7%, which is much higher than the contribution of monoatomic Se vacancies, 37.7%, and even higher than the contribution of diatomic Se vacancies, 40.4%.

The impact of the location of the defects on the thermal conductivity has also been investigated. Comparing the thermal conductivity of SS-c and SS-e, it is learnt that at 300 K, the thermal conductivity of SS-e is 42.1 W/(m·K), while that of SS-c is 41.2 W/(m·K). The location of the defect has an important influence on the efficiency of heat transport efficiency [34, 35]. The thermal conductivity of SS-e is slightly higher than that of SS-c. It indicates that the Se atoms in the central place contribute more to heat conduction than that at the edge location. Simultaneously, contrasting the thermal conductivity of DS-s and DS-d, it is discovered that at 300 K, the thermal conductivity of DS-s is 39.4 W/(m·K), and the thermal conductivity of DS-d is 38.4 W/(m·K). Defects located in the same layer contribute slightly less to the reduction of thermal conductivity than defects located in different layers. It shows that scattered defects facilitate more to reduce thermal conductivity than comparatively concentrated defects.

By comparing the termwise decrease ratio, as shown in Figure 6, it is found that at 110 K, the thermal conductivity decreases by 12.8%, 18.0%, 17.8%, 14.8%, 18.1%, 17.7%, respectively for PS, SS-c, DS-s, SW-c, SS-e, DS-d, respectively. Structures without defects have the smallest decline ratio, while structures with defects have a relatively large ratio. As the temperature increases, the decreasing ratio gradually stabilizes, and the difference with the defect-free structure is getting smaller and smaller, but the overall ratio is still higher than that of the defect-free structure. Therefore, the defective structure can not only block the phonon heat transfer path and reduce the thermal conductivity, but also be more sensitive to temperature changes. A small temperature change can bring about a greater decrease in thermal conductivity.

One possible reason for the large thermal conductivity drop caused by defects is that the existence of vacancies causes the lifetime (or the mean free path) of certain phonon modes to decrease [36-38]. Vacancies will scatter the phonons strongly. The greater the density of the vacancies and the larger the size of the vacancies, the greater the impact. Some acoustic vibration modes will be path-blocked and localized, thereby adversely affecting the thermal conductivity. The influence of vacancies on the thermal conductivity is not linear, but as the number of vacancies increases, the influence on the decrease in thermal conductivity is getting smaller and smaller. The effect of the Se vacancy on the mean free path is far less sensitive than the W vacancy.

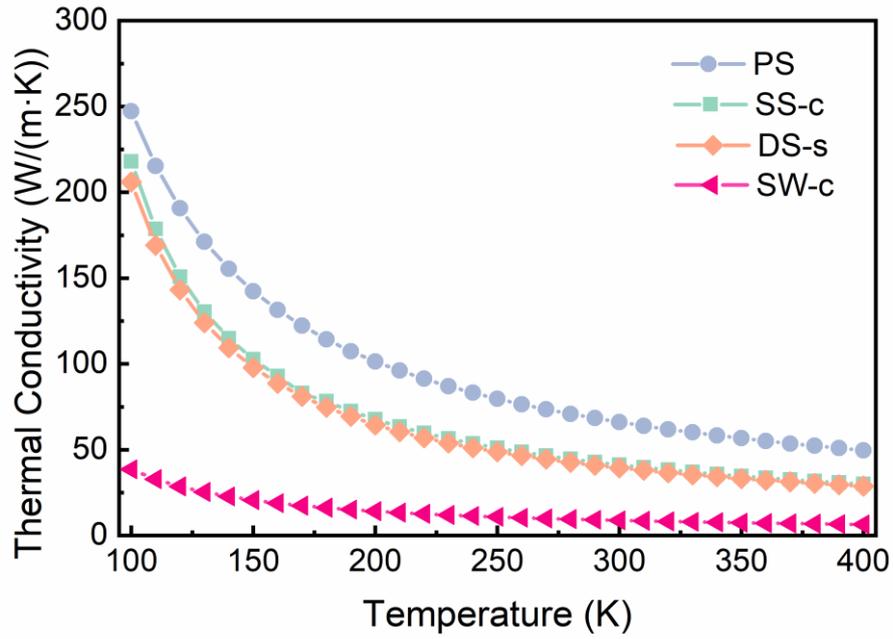

Fig 4. Thermal conductivities of WSe$_2$ under different temperatures of PS, SS-c, DS-s, SW-c.

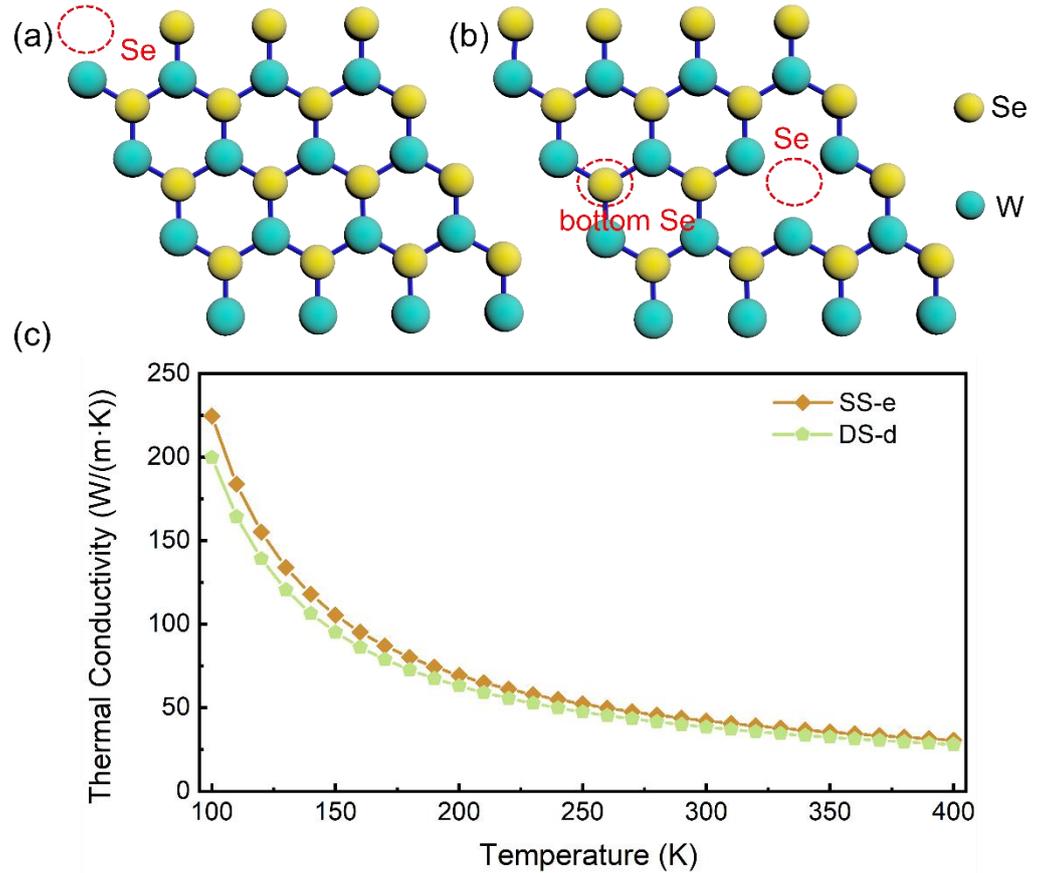

Figure 5. Schematic diagram of (a) SS-e, (b) DS-d. (c) Thermal conductivities of WSe$_2$ under different temperatures of SS-e and DS-d.

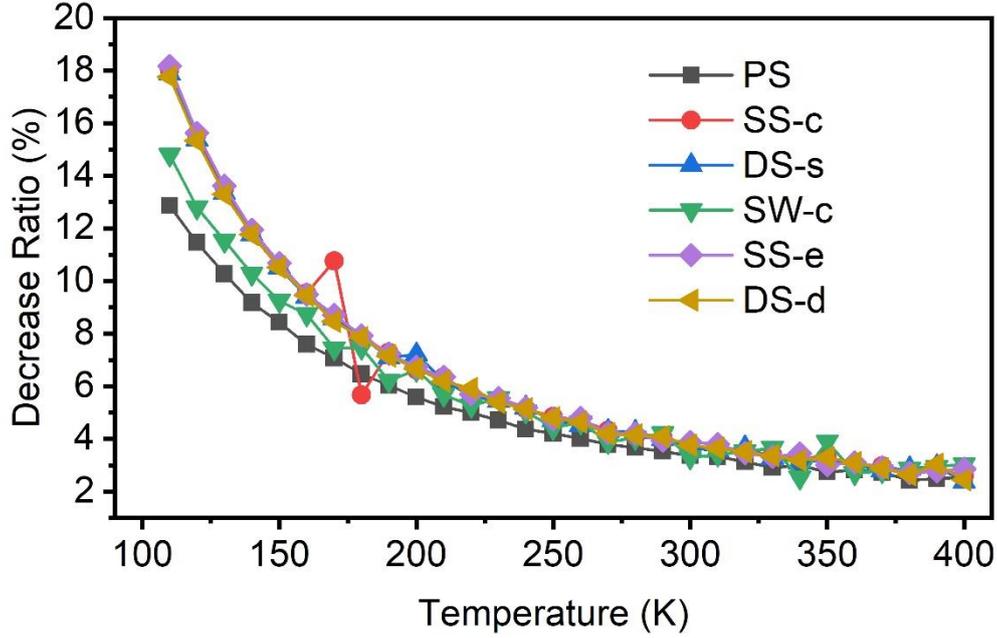

Figure 6. Temperature dependent thermal conductivity decrease ratio.

Table 1. Thermal conductivities of different defective WSe$_2$.

| T (K) | PS | SS-c | DS-s | SW-c | SS-e | DS-d | T (K) | PS | SS-c | DS-s | SW-c | SS-e | DS-d |
|---|---|---|---|---|---|---|---|---|---|---|---|---|---|
| **100** | 247.2 | 217.9 | 205.9 | 38.5 | 224.5 | 199.8 | **260** | 76.5 | 48.6 | 46.5 | 10.3 | 49.7 | 45.3 |
| **110** | 215.4 | 178.6 | 169.1 | 32.8 | 183.7 | 164.3 | **270** | 73.6 | 46.5 | 44.5 | 9.9 | 47.6 | 43.4 |
| **120** | 190.7 | 150.8 | 143.1 | 28.6 | 155.0 | 139.1 | **280** | 70.9 | 44.6 | 42.6 | 9.5 | 45.6 | 41.6 |
| **130** | 171.1 | 130.4 | 124 | 25.3 | 133.9 | 120.6 | **290** | 68.4 | 42.8 | 40.9 | 9.1 | 43.8 | 39.9 |
| **140** | 155.4 | 114.9 | 109.4 | 22.7 | 117.9 | 106.4 | **300** | 66.1 | 41.2 | 39.4 | 8.8 | 42.1 | 38.4 |
| **150** | 142.3 | 102.7 | 97.9 | 20.6 | 105.3 | 95.2 | **310** | 63.9 | 39.7 | 38 | 8.5 | 40.5 | 37 |
| **160** | 131.5 | 92.9 | 88.7 | 18.8 | 95.3 | 86.2 | **320** | 61.9 | 38.3 | 36.6 | 8.2 | 39.1 | 35.7 |
| **170** | 122.2 | 82.9 | 81.1 | 17.4 | 87.0 | 78.9 | **330** | 60.1 | 37 | 35.4 | 7.9 | 37.8 | 34.5 |
| **180** | 114.3 | 78.2 | 74.7 | 16.1 | 80.1 | 72.7 | **340** | 58.3 | 35.8 | 34.3 | 7.7 | 36.5 | 33.4 |
| **190** | 107.4 | 72.5 | 69.4 | 15.1 | 74.3 | 67.5 | **350** | 56.7 | 34.6 | 33.2 | 7.4 | 35.4 | 32.3 |
| **200** | 101.4 | 67.7 | 64.4 | 14.1 | 69.3 | 63 | **360** | 55.1 | 33.6 | 32.2 | 7.2 | 34.3 | 31.3 |
| **210** | 96.1 | 63.4 | 60.4 | 13.3 | 64.9 | 59.1 | **370** | 53.6 | 32.6 | 31.3 | 7.0 | 33.3 | 30.4 |
| **220** | 91.3 | 59.8 | 57 | 12.6 | 61.2 | 55.6 | **380** | 52.3 | 31.7 | 30.4 | 6.8 | 32.4 | 29.6 |
| **230** | 87 | 56.5 | 53.9 | 11.9 | 57.8 | 52.6 | **390** | 51 | 30.8 | 29.5 | 6.6 | 31.5 | 28.7 |
| **240** | 83.2 | 53.6 | 51.1 | 11.3 | 54.8 | 49.9 | **400** | 49.7 | 30 | 28.8 | 6.4 | 30.6 | 28 |
| **250** | 79.7 | 51 | 48.7 | 10.8 | 52.2 | 47.5 | | | | | | | |

Unit: W/(m·K)

## 4. Conclusion

The thermal properties of monolayer WSe$_2$ with several kinds of defect structures have been studied by first-principles calculations. By solving Boltzmann transport equations implemented in ShengBTE package via inputting the second-order harmonic and third-order anharmonic IFCs, the temperature-dependent thermal conductivities of WSe$_2$ of defect structures can be calculated. The thermal conductivity of perfect-structure WSe$_2$ at 300 K is 66.1 W/(m·K), while the defect samples have values of 41.2, 39.4, 8.8, 42.1, and 38.4 W/(m·K), respectively for SS-c, DS-s, SW-c, SS-e, and DS-d. Sample with defects have much lower thermal conductivity compared with perfect structure, up to 86.7% decrease (SW-c). More defect content can lead to lower thermal conductivity by comparing PS, SS-c, DS-s samples. Also, a single W atom defect create a quite lower thermal conductivity even than double Se atom defect structure by the comparison of PS, SS-c, and SW-c. Finally, the impact of defect location on thermal conductivity has been explained by comparing SS-c and SS-e, DS-s and DS-d. the Se atoms in the central area contribute more to heat conduction than that at the edge place. Also, scattered defects facilitate more to reduce thermal conductivity than comparatively concentrated defects. By introducing atom defects, we can reduce and regulate the thermal property of $WSe_2$.

**Acknowledgements**

The authors are grateful for the computation resources provided by Pittsburgh Supercomputing Center and Argonne Leadership Computing Facility.